\documentstyle[12pt]{article}

\begin{document}

\author{Gary R. Goldstein\thanks{To be published in Peace and Change: a
Journal of Peace Research (2000). Submitted 13 March 2000.}\\
Department of Physics\\
Tufts University\\
Medford, MA 02155 USA}
\title{A Review Essay: ``Lise Meitner and the Dawn of the Nuclear Age'',
by Patricia Rife}
\maketitle

\begin{abstract}
The recently published book, ``Lise Meitner and the Dawn of the Nuclear
Age'', by Patricia Rife (Boston: Birkh\"{a}user, 1999) is reviewed
in an essay for the lay audience. Meitner was a leading nuclear physicist
at the time that the nucleus was the most exciting frontier of
science. To establish her career, she had to overcome daunting prejudices
against women in science and academia. Being of Jewish origin in Germany
in the 1930's, she narrowly escaped certain disaster. Meitner was a
crucial participant in the discovery of nuclear fission, yet did not share
in the Nobel Prize that her collaborator, Otto Hahn, received in 1945. How
these events came about, how they were intertwined with contemporary
history and how they fit into the evolution of Meitner's social
conscience and her abhorrence of war are some of the fascinating subjects
discussed in the book and reviewed in this essay.
\end{abstract}

\newpage

Science has contributed immeasurably to the history and character of the 
twentieth century. Advances in the fundamental understanding of nature 
raced along at a mindboggling pace. The speedy applications of those 
advances have led to everything from palm size computers and wireless 
telephones to microwave ovens and nuclear weapons.  Consequently, one of 
the important extra-scientific questions for any scientist has become ``To 
what extent are you responsible for the applications of your work?'' And 
concomitantly, to what extent should the scientist use his or her 
expertise and reputation to influence the social uses of their science. 
We need only think about the many innovations in military technology 
during the two World Wars to realize how important these questions are. 
Even the most narrowly focused practitioner of ``pure  research'' can be 
faced with these questions. Consider Albert Einstein, a pacifist, whose 
extraordinary insights into the nature of space and time ultimately made 
possible the building of the A-bomb, the most destructive single weapon 
ever used in war.  He became an outspoken critic of American nuclear 
policy after the war, and was labeled a ``fellow traveler'' by some of the 
cold warriors.

Lise Meitner, a contemporary of Einstein's, was a remarkable nuclear 
physicist whose discovery of nuclear fission paved the way for the 
Manhattan project, although she was unaware of the project itself.  She 
did not share in the credit for that discovery, in any case, having been 
passed up by the Nobel Prize committee, while her collaborator, Otto 
Hahn, did receive the prize in 1945. How these circumstances came about, 
and how they fit into the evolution of her social conscience and her 
abhorrence of war are some of the fascinating subjects discussed in the 
biography by Patricia Rife, ``Lise Meitner and the Dawn of the Nuclear
Age'',~\cite{rife} here under review.

Lise Meitner's name was hardly known in recent times, until an earlier 
biography by Ruth Lewin Sime~\cite{sime} appeared in 1996 that reminded
readers -- interested scientists, primarily -- of her importance. Yet, in
the aftermath of World War II, Meitner toured the US and was acclaimed ``the 
Mother of the A-bomb'' in the popular press (a dubious honor that she 
shunned). As that war and its memories have faded from public 
consciousness, so have the names in a long list of scientists and 
engineers whose work contributed to weapons research in general and the 
development of the nuclear fission bomb in particular.  Certainly, 
everyone recognizes Albert Einstein, whose role in the dawn of the 
nuclear age was crucial, although he too had nothing to do with the work 
of bomb project. He is now Time Magazine's ``Person of the Century'' and a 
logo for consumer products. Ironically, even among the cognoscenti, 
contemporary scientists for whom Einstein's work is part of their life's 
breath, Meitner's essential role in nuclear research is hardly known. Nor 
are the details of her extraordinary life. Her devotion to scientific 
discovery was realized amidst an interminable struggle against misogyny, 
anti-Semitism, economic hardship and indifference. Patricia Rife's 
biography, which was begun 11 years earlier, further contributes to the 
understanding of the struggle and achievements of Lise Meitner.

Some of the important milestones discussed by Rife are summarized herein. 
Lise Meitner was born in 1878 of a middle class Jewish family in Vienna. 
Early on it was clear that she had special ability in science and 
mathematics, although it was very difficult for a young woman to obtain a 
suitable education. She persisted, struggled against an unsympathetic 
system and managed to be the first woman ever admitted to the physics 
department at the University of Vienna. Her subsequent career was 
exceptional. From her twenties through her sixties she pursued the most 
vexing and interesting puzzles in nuclear and quantum physics. And she 
unraveled many of those puzzles in her research.

The times and places in which Meitner lived and worked coincided with 
some of the century's most important and terrifying events. She was among 
the first women in Austria and then Germany to be allowed participation 
in higher education, both as a student, a researcher and finally as a 
professor. At the time she completed her Ph.D. in Vienna, many of the new 
discoveries in microscopic physics were being made and discussed. The 
nature of the quantum world was beginning to be explored, and it was 
becoming clear that the laws of physics were in need of radical change in 
order to account for the new phenomena. In 1907, already involved in 
studies of radioactive material, Lise decided to move to Berlin where Max 
Planck was leading a group of young physicists and chemists in the first 
forays into the quantum world. Within a few years some of the leading 
proponents of the new physics would be there - Einstein, von Baeyer, 
Franck, Hertz, Stern, von Laue. Lise soon began to collaborate with a 
young chemist whose interests were similar to hers, Otto Hahn, in a 
partnership that would last through most of her career. But for many 
years she was an unpaid participant in the research at the newly 
constituted Kaiser Wihelm Institute for Chemistry, women not being 
allowed any official status. 

Science is not advanced in a vacuum of course, and Rife fills in the 
historical context throughout. Although Meitner was incredibly focused on 
her nuclear studies, she was forced to be aware of the powerful misogyny 
that dogged her most pointedly in the early part of her career, but 
continued in many forms throughout her life. The turmoil of war - World 
War I - and its sweep of scientific talent into weapon making or 
soldiering amidst injuries and death were a sobering experience. She 
eventually volunteered, seemingly reluctantly, to aid the Austrian war 
effort as a nurse on the frontlines. She was asked to use her expertise 
in x-ray properties to aid the diagnosis of severe battlefield injuries 
and to train others in the fledgling medical technology. She worked to 
exhaustion.

These war experiences did not turn her into an overt pacifist, however 
(in a letter to Hahn she complained about Einstein's pacifism being out 
of touch with the reality of the war engulfing Europe), but left her 
deeply skeptical about political motives for committing lives to 
furthering grandiose nationalistic goals. She was very unhappy about her 
peers' participation in research on poison gas, which was led by the 
distinguished chemist Fritz Haber. Otto Hahn even took part in one of the 
raids that used gas on the Allies' troops, for which Meitner chastised 
him. In the midst of all this turmoil she longed for the peaceful pursuit 
of her research, and returned to Berlin hoping for the insanity to ``pass 
over''. It eventually did, but was replaced by the post-war deprivations 
that made everyday life a struggle. Rife emphasizes the point that Lise 
developed an expectation that in time political and economic chaos would 
run their course. The best thing to do in the meantime was to focus on 
her work. One of the people she admired most in Berlin, Max Planck, who 
was also a friend and mentor, advocated the attitude of the scientific 
internationalists, that science transcends politics and nations. It 
should be pursued for its own sake. Lise, following that philosophy, 
maintained the work of the Kaiser Wilhelm Institute for Chemistry 
Radiological Studies group as the war came to an end and she continued on 
as the director of the Physics Section. Her achievements multiplied and 
her reputation grew among the increasing number of scientists devoted to 
nuclear research.

Meitner finally had been given an appointment, the first for a woman, in 
the University of Berlin physics department (a separate and more 
internationally prestigious institution than the Kaiser Wilhelm 
Institute) in 1914, but not with a salary commensurate with her status. 
For many years she had been getting by with help from her family and a 
stipend for being Planck's assistant, while working at the Institute. 
Finally in 1916, at the age of 38, she was given a salary nearly equal to 
Hahn's, having been his equal in responsibilities for several years 
already. As the post-war years were succeeded by the Weimar Republic 
period, her research progressed unhampered by the forces of political 
change. But those forces became overwhelming with the collapse of the 
liberal Weimar government.

The work on nuclear fission, for which Lise Meitner will be remembered 
most, has a complicated history that is enmeshed with the horrendous 
policies of Nazi Germany. Clearly presenting the history of the discovery 
and the ensuing establishment of priority, within the tumultuous events 
of the times, is a major achievement of the Rife biography. This slice of 
modern history sharply illustrates how inescapable were the tentacles of 
fascism, even among the most distinguished scientists. And the permanent 
dislocation of one individual's hitherto exemplary and productive life is 
a microcosm of the disruptions of the time. What is fascinating from the 
perspective of the history of science, is how inextricably connected are 
the demands of politics and the military with the directions taken by 
scientific research. Once fission was identified by Meitner and 
colleagues, an international race was underway to exploit its military 
potential. To summarize the intertwined issues that Rife disentangles is 
difficult, but essential for appreciating the complexity of the task Rife 
has undertaken.

To set the stage, in 1933 Meitner's professorship at the university was 
revoked by the Nazi directorship, to her great shock. Nevertheless, she 
managed to remain at her beloved Kaiser Wilhelm Institute for Chemistry 
for five more years during the continuing rise of the fascists and the 
imposition of their anti-Jewish laws. At the same time as the revocation, 
Planck, Heisenberg and von Laue nominated Meitner and Hahn for the Nobel 
Prize for their pioneering work in radiochemistry. These nominations were 
repeated over many years. It was hoped that as a side benefit, a Nobel 
Prize would ease the plight of the non-Aryan Meitner and legitimize her 
remaining in her research position at the Institute, but to no avail for 
her. Relying naively on her experience during World War I, Lise tried to 
``hold center'', continuing with her work, while political chaos and the 
deplorable attitudes of many new colleagues (already by 1933 half of the 
institute staff were Nazi party members) began to engulf her and her 
colleagues. In protest to the dismissal of Jews from the university, Hahn 
resigned his university position (but not his institute position, which 
was more significant as the university lapsed into fascist polemics). 
However, Planck withdrew from the fray after an initial protest to Hitler 
over the dismissal of Fritz Haber from the directorship of the Kaiser 
Wilhelm Institute for Chemistry. Other scientists in Lise's circle 
advised avoiding direct confrontation with the Nazis. Rife emphasizes 
that Otto Hahn's choice of ``compromise rather than righteous 
indignation''(p.124 in \cite{rife}) was symptomatic of many educated
German's 
unwillingness to confront the Nazi hierarchy.

Rife next carefully reconstructs the chronology of the nuclear research. 
While getting the facts straight, Rife is quite terse with the important 
scientific developments. A more readable and conventional account of the 
science, without the detailed chronology of Meitner's contribution, can 
be found in the excellent book, ``The Making of the Atomic Bomb'', by 
Rhodes~\cite{rhodes}. The essentials follow.

In 1934, in the midst of the very threatening political environment 
throughout Europe, Enrico Fermi and his research group in Rome initiated 
a new line of nuclear investigations in which neutrons were used to 
bombard a variety of different nuclei and the byproducts of the 
interactions were studied. The transmutation of the elements was 
underway. Lise and Otto Hahn resumed their collaboration after a hiatus 
of 12 years in order to combine their skills in pursuing this new and 
promising approach. Their ``Berlin Group'', which added Fritz Strassmann, 
was immediately a major contributor to the burgeoning research area, 
competing with Fermi's group and the ``Paris Group'' of Irene Curie and 
colleagues, along with several newly constituted US teams. The early 
results of the Berlin group suggested to Meitner that ``transuranic 
elements'', artificially created nuclei heavier than the naturally 
occurring heaviest element uranium, were sometimes produced when neutrons 
hit heavy nuclei. By observing the beta radiation from the radioactive 
nuclear byproducts, Meitner was able to identify many of those nuclei. 
Hahn and Strassmann, both chemists, did the chemical analyses of the same 
byproducts. But there were some puzzling discrepancies with the results 
of the Curie experiments. Thinking that they were observing transuranic 
nuclei, Meitner couldn't accept that the French were finding a lighter 
element, lanthanum, among the nuclei. Furthermore, issues of priority 
developed with Curie's group concerning the transuranics. These were very 
exciting times for science. A new area of exploration, the physics of the 
nucleus, was yielding new discoveries at a rapid rate and attracting 
considerable attention. Competition was keen; major discoveries were 
imminent. There are few periods in the history of modern physics and 
chemistry that were as heated. Hahn and Meitner planned a series of 
experiments to probe further into the byproducts that result when uranium 
is the bombarded nucleus. But these experiments were slowed and 
interrupted by the events of 1938, as Rife describes. 

The Anschluss, the German takeover of Austria in March 1938, put all 
Austrians under the German laws and Lise Meitner was classified as a Jew. 
Rife doesn't note that Meitner had actually converted to Protestantism in 
her 20's (Sime discussed this in her book~\cite{sime}), although this was 
irrelevant for the ``racial'' distinctions that the Nazis made. It was 
clear to Meitner and her friends, inside and outside of Germany, that she 
had to leave as soon as possible; ``holding center'' was no longer viable. 
Hahn's efforts to allow her to maintain her position at the Institute 
were half-hearted and ineffectual. She was quite angry with him for not 
speaking out more forcefully on her behalf. 

The subsequent fits and starts arranging Meitner's escape constitute an 
adventure story that Rife tells well. In brief, Lise managed, with no 
proper exit visa, but with extraordinary help from the Dutch physicists 
Coster and Fokker, to get to the Netherlands, where she remained a short 
time before going to Bohr's institute in Copenhagen and then Stockholm, 
Sweden. Sweden offered her a modest position in a newly constituted 
institute for nuclear research. Unfortunately the head, Nobel Laureate 
Manne Siegbahn, was never very interested in Meitner's work and allowed 
her only quite inadequate facilities. The Sime book~\cite{sime} is quite
damning 
of Siegbahn's possible motivation for so ignoring Lise's research needs, 
while Rife puts more emphasis on his preoccupation with building big 
machines. In any case, Meitner's research in Sweden was hampered by 
inadequate equipment and lack of assistants. She was quite frustrated and 
probably quite depressed, missing her beloved physics research, the 
Berlin institute, colleagues, friends and even her native German 
language. 

So from July through September 1938 Meitner could do no research, but 
Hahn and Strassmann continued the project in Berlin to determine the 
byproducts from the bombardment of uranium by neutrons. Hahn frequently 
wrote to Meitner about their progress and asked for her advice and 
interpretation. On Dec. 16 Hahn and Strassmann, by careful measurements, 
had convinced themselves that one of the byproducts was definitely 
barium, an element of slightly more than half the mass of the uranium. 
Hahn wrote Lise asking her to interpret this puzzling result, realizing 
that this could not be the signature of one of the transuranics. 
Reluctant to give up the transuranic idea, Lise first tried to find some 
loophole in the results. While letters were being exchanged, Hahn wrote 
the article about the work without consulting Strassmann and without 
Meitner's name as a co-author. In the meantime Lise spent the Christmas 
holiday period in western Sweden with friends and her nephew, the young, 
recently emigrated physicist Otto Frisch. The two, aunt and nephew, 
pondered the meaning of the barium results and, during a walk in the 
forest, had a breakthrough in understanding. Frisch recalled later that, 
in classic fashion, Lise sat down by a tree and started doing rough 
calculations on a scrap of paper. They had realized that the barium was 
appearing as a result of the uranium actually splitting into smaller 
nuclei after being struck by a neutron. And the amount of energy released 
was just what Einstein's famous formula, E = mc$^2$, predicted. Fission
was discovered! 

It is difficult to imagine the emotional turmoil Meitner experienced in 
that short time from the urgent arrangements being made for her escape 
from Germany in April 1938 to the quiet Christmas holiday in western 
Sweden. She had been engaged in some of the most exciting research of her 
career, while trying to keep the terrifying external situation for 
non-Aryans away from consciousness. Then in March she was forced to 
confront the fact that her life was in grave danger.  The time for 
legally obtaining an exit visa had passed. The escape plans were fraught 
with dangerous uncertainty. She survived the fearful anxiety of crossing 
the border to safety, only to encounter further uncertainty about her 
livelihood in foreign lands. What distinguishes these emotional upheavals 
from so many other refugees' experiences is the wrenching removal from 
her life's work, work that was on a threshold of momentous discovery.  
She then had to endure reading Hahn's letters about the continuation of 
that research, while she was unable to do any work herself - she was 
experimentally mute. But then, when she and Frisch explained the 
experimental results in terms of fission, a term they invented, she was 
at the height of scientific creativity again. And at the age of 60, this 
creative insight was quite remarkable. Nevertheless, Lise was again in 
the doldrums after the holiday when she returned to the isolation of the 
primitive lab she had been assembling in Stockholm. Surely she had to 
have been depressed, as Rife surmises. A superhuman fortitude sustained 
her through the war years; she managed to do some interesting research in 
spite of the obstacles.

After Lise wrote Hahn about the physical meaning of the experimental 
results, she and Frisch composed an article whose submission and 
publication were quite slow. Long before the article appeared in print a 
remarkable set of circumstances, involving Niels Bohr, brought the news 
of fission to the US and a race to study the fission process was 
underway. Meitner was out of the loop from then until the end of World 
War II. Her experiments conducted in that period were no longer at the 
cutting edge.  The real developments of nuclear fission became part of 
the secret rush to build a nuclear bomb by the US, Germany and later, the 
Soviet Union. 

There are many questions raised by this history of fission's discovery. 
Why was Meitner not included as a co-author on Hahn and Strassmann's 
paper? Rife doesn't consider this point, but the fact that Meitner had 
been an active partner in this particular research until her escape in 
July reasonably should have led to her inclusion on the paper. Sime's 
book~\cite{sime} speculated that it might have been dangerous for Hahn to
include 
this recently escaped non-Aryan as a co-author. Most of the scientists 
involved in this line of research knew that Meitner had been an equal 
partner in the work anyway. Unfortunately, though, after the war Hahn 
continued to downplay Meitner's role in the research as Rife shows. 
During the war ``He refused to take a stand on the politics in and out of 
his institute: the Third Reich was blinding Hahn and he began to discount 
Meitner's insights and contributions he had frantically sought out months 
earlier. We witness here appeasement, professional cowardice, and worse.''
(p.213). Perhaps Hahn's initial lapse is understandable given that the 
groundbreaking paper may have been rejected by the editor or some other 
authorities for political reasons, although Hahn knew the editor 
personally. More likely, he would have had to stand up to the Nazi 
directorship of the Institute and the official scientific establishment, 
thereby jeopardizing his position. And when he finally received 
recognition for the paper and the acclaim from his German colleagues that 
followed, he was not about to share credit with Meitner. There is no 
question that his behavior was scurrilous. He must have constructed a 
torturous self-justification to assuage his guilt for deserting his 
decades long collaborator and friend. He never adequately acknowledged 
her essential role in the discovery, even long after he was awarded the 
Nobel Prize and was a leading figure in post-war German science.

Another important element of the discovery of nuclear fission, for 
historians of science, is how Meitner and Hahn were able to accept the 
error in their previous thinking about their transmutation research. Rife 
notes that both shared an initial unwillingness to abandon their 
interpretation of their previous neutron bombardment work. Here was an 
archetypal struggle at the beginning of a paradigm shift~\cite{kuhn} or
creative 
leap. Hahn remained unsure of his experimental results for a while, but 
once Meitner, with Frisch, realized they had seen fission in the data, 
everything fell into place.

The war years were difficult for Meitner for many reasons. The news of 
horrendous destruction and turmoil was depressing. She helped friends 
fleeing Nazi persecution. She kept in touch with family and was able to 
communicate with some of her Berlin colleagues, but must have felt quite 
helpless in the sweep of events. Her nephew Frisch had managed to get a 
position in Birmingham, England and was continuing to experiment with 
fission. In collaboration with another German-Jewish émmigrant, Rudolph 
Peierls, he figured out how the fission of a rare isotope of uranium 
(U235) could be used to initiate a ``chain reaction'' and create a ``super 
bomb'' of enormous power. The secret Frisch-Peierls
memorandum~\cite{frisch} became 
the impetus for the start of the Manhattan Project. Unbeknownst to Lise, 
her nephew, Peierls and other physicists working in Britain (including 
Klaus Fuchs) were sent off to Los Alamos, New Mexico to contribute to the 
making of the bomb. Had she been given the opportunity to participate in 
that war work, there is little doubt that Meitner would have declined, as 
her previous history suggests. In a separate discussion, Rife makes the 
interesting point that the scale of scientific research would never be 
the same after the ``big science'' projects became the rule with the 
infusion of large government budgets into scientifically innovative 
weapons development. Siegbahn's institute in Stockholm had become an 
example of big science that left Meitner out in the cold (although not 
directed toward weapons work).

When the war in Europe ended, Meitner was horrified to learn about the 
deaths and deprivations of the millions of victims of the Nazi 
concentration camps. She could not forgive her German colleagues for 
their lack of active opposition to the regime. Rife quotes a remarkable 
letter to Hahn (that he never received) in which Meitner excoriates her 
old friend and research partner: ``All of you lost your standards of 
justice and fairness. ... All of you also worked for Nazi Germany, and 
never even attempted passive resistance. Of course, to save your troubled 
conscience, you occasionally helped an oppressed person; still, you let 
millions of innocent people be murdered, and there was never a sound of 
protest.''(p.249). Later she singles out Heisenberg, ``They should force a 
man like Heisenberg, and millions of others with him, to see these camps 
and the tortured people.''(p.250). Heisenberg had come to Bohr's institute 
in Copenhagen in 1941 and had delivered a lecture full of propaganda for 
the regime that had infuriated Meitner. If there remain any doubts about 
Heisenberg's lack of scruples about his leading Nazi nuclear bomb 
research, Lise would not have shared them.

Bohr's efforts to assure the establishment of Lise's priority for the 
interpretation of fission were not significant enough to prevent the 1944 
Nobel Prize in Chemistry from being awarded (in 1945, after the War) to 
Otto Hahn exclusively for the discovery of nuclear fission. ``Lise Meitner 
keenly felt the injustice of the situation, as did many of her 
colleagues'', Rife recounts (p.258). In his Nobel address to the Royal 
Academy of Sciences, with Lise present, Hahn gives credit to Meitner and 
Frisch for interpreting his and Strassmann's results. Nevertheless, Hahn 
did not, in turn, nominate her for the prize once he was on the 
nominating committee. He gave her part of the monetary award, though, 
which she then sent on to an Emergency Committee of Atomic Scientists 
chaired by Einstein in Princeton.

At war's end Lise received unexpected attention from the press. She was 
sought out after the atomic bombing of Japan. Called the ``mother of the 
A-bomb'', she quite forcefully tried to make it clear that she was 
completely removed from any weapon research. Rife quotes a friend of hers 
who recalls that when Lise first heard about Hiroshima, there were ``tears 
- shock - and then silence''(p.252). With the attention came some welcome 
invitations for her to visit the US. She was asked to give lectures at 
Catholic University in Washington, DC, and was a guest of honor at the 
Women's National Press Club. She traveled to many universities giving 
lectures and being honored. She spoke more and more about the importance 
of encouraging women to enter higher education and scientific research. 
She also found herself among a growing number of scientists who were 
profoundly worried about the nuclear Pandora's box that had been opened.

Just days after the bombing of Hiroshima, Lise Meitner took part in a 
radio interview conducted by Eleanor Roosevelt in which she said: ``Women 
have a great responsibility and they are obliged to try, so far as they 
can, to prevent another war. I hope that the construction of the atom 
bomb not only will help to finish this awful war, but that we will be 
able to also use this great energy that has been released for peaceful 
work.''(p.253). Here she is expressing the attitudes that defined the 
early post-war anti-nuclear movement among the atomic scientists. It has 
to be said that history shows, however, that it is doubtful that the atom 
bomb helped to finish the war or made it less awful~\cite{rhodes}. And the
hope that 
atomic energy should be harnessed for peaceful work has been quixotic, 
due to the enduring safety problems of nuclear reactors. Nonetheless, her 
words were quite courageous at that time. The plea to women is certainly 
heartfelt, given her own history. Lise felt that even she had abdicated 
her moral responsibility by staying in Germany as long as she had, in 
spite of the personal danger, because her work gave legitimacy to the 
regime.  She urged women and scientists to be more aware of the moral 
consequences of events around them.

Meitner spent the post-war years traveling, giving lectures, advocating 
arms control and the equal participation of women in science. She wanted 
to share the lessons that she learned in a lifetime of struggle against 
the ravages of war and the prejudices toward women and Jews. She lived an 
active and dedicated life to the age of 89.

Although Rife gives us a thorough accounting of the facts of Meitner's 
life, and indicates the concerns she had for family and friends 
throughout, there is a missing personal connection. For most of Lise's 
years she lived alone, but enjoyed the company of friends. Did she 
consciously decide to have no intimate romantic relationships? Did the 
battle to be accepted in a man's world preclude forming such attachments? 
Or have the biographers missed something? Did physics and her recreations 
- listening to music and walking - fill her time?  We can only infer that 
she was depressed at particular times when circumstances were quite 
difficult. What is the core of her personality? Lise was modest and very 
shy in public, yet unstoppably competitive and tenacious in her work. She 
published papers at a high rate, even by today's standards (when the 
competition for jobs and advancement is unpleasantly keen). Was this her 
strategy to overcome the prejudices against women in science? Eventually, 
with all of her achievements, she was accepted by the mostly male 
community of nuclear scientists. Though she was never given the same 
level of recognition and reward as her collaborator Hahn, Lise Meitner's 
single-minded dedication to physics while faced with a myriad of 
obstacles is proof of an extraordinary person. And with such 
extraordinary people we want to know how they came to be. We are 
ultimately left to theorize for ourselves, but Rife has given us much of 
the data we need in a well written, thorough, readable and engrossing 
work. The book is clearly a paean to a great woman scientist.

\end{document}